\documentclass[superscriptaddress,twocolumn,preprintnumbers,amsmath,amssymb,showkeys]{revtex4}

\usepackage{graphicx}
\usepackage{pdfpages}
\usepackage{dcolumn}
\usepackage{bm}

\begin{document}

\texttt{preprint}

\title{The Confinement of Vortices in Nano-Superconducting Devices}

\author{W.M. Wu}
\affiliation{
Department of Physics, Loughborough University, Loughborough LE11 3TU,
United Kingdom}
\author{M.B. Sobnack}
\affiliation{
Department of Physics, Loughborough University, Loughborough LE11 3TU,
United Kingdom}
\author{D.M. Forrester}
\affiliation{Department of Chemical Engineering, Loughborough University,
Loughborough LE11 3TU,
United Kingdom}
\author{F.V. Kusmartsev}
\affiliation{
Department of Physics, Loughborough University, Loughborough LE11 3TU,
United Kingdom}

\date{\today}

\begin{abstract}
We have investigated the confinement of 3-D vortices in specific cases of Type-II ($\kappa = 2$) nano-superconducting devices. The emergent pattern of vortices greatly depends on the orientation of an applied magnetic field (transverse or longitudinal), and the size of the devices (a few coherence lengths $\xi$). Herein, cylindrical geometries are examined. The surface barriers become very significant in these nano-systems, and hence the characteristics of the vortices become highly sensitive to the shape of the system and direction of an applied field. It is observed that nano-cylindrical superconductors, depending on their sizes, can display either first or second order phase transitions, under the influence of a longitudinal field. In the confined geometries, nucleation of a giant vortex state composed of a n-quanta emerges for the longitudinal magnetic field.  

\end{abstract}


\maketitle

\section{Introduction}
Generally, there are two types of superconductors: Type-I and Type-II. In macroscopic sized Type I superconductors in an applied magnetic field there emerges a perfect diamagnetic state, whereby the field is completely expelled from the sample. This expulsion is known as the Meissner-Oschenfeld effect. In strong magnetic fields, the superconducting state vanishes, and a normal metallic one is restored \cite{Lond1,Lond2}. However, another category of superconducting materials, Type II, can react quite differently to an applied field. In weak fields, these superconductors behave identically to the Type I systems by completely expelling the field. But as the strength of the magnetic field increases, penetration into the sample occurs to create flux tubes of normal conductivity surrounded by spiralling superconducting currents. These small tornadoes of superconducting current are called vortices. The Type II superconductor allows the nucleation of quantised vortices in a triangular pattern (the Abrikosov lattice). However, these facts are only going to be true in the bulk superconductor, but not in the nano-sized superconductors. In fact, the vortices and superconductors themselves, could exhibit abnormal behaviours in some confined nano-geometries \cite{Dorsey}. Recent experiments \cite{Zhang,Michotte,Stenuit,Bend1,Bend2} have indicated that Type I (Pb) superconducting cylinders may have some Type II characteristics, possibly suggesting that vortices may quite likely nucleate in Type I superconductors with restricted nano-geometries \cite{Zhang,Geim2,Aliev,Gasparovic,Ishii}. 
In a restricted region, it is not necessary for a vortex to be a quantised fluxoid, as proposed by Bardeen \cite{Bardeen} and observed by Geim {\it et al.} \cite{Geim1}, because of the flux spilling out of the superconducting surface \cite{Geim3}. It is also found that nano-superconductors could undergo the $1^{\text {st}}$ or $2^{\text {nd}}$ order phase transitions, depending on the size and surface barrier of the system \cite{Geim1,Moshchalkov}. Some reports further showed that superconducting cylinders suffer a stronger Meissner-Ochsenfeld effect in a transverse field than in a longitudinal field, and concluded that the magnetisation response is dependent on the direction of an applied magnetic field \cite{Zhang}. 

The surface barriers in nano-superconductors play a very important role to influence the interaction of vortices \cite{Bean,Novoselov,Peeter1}. Vortices are sensitive to the nano-boundaries \cite{Strongin,Thomas}, experiencing an attractive force from the boundaries and a repulsive force due to the magnetic field \cite{Bean}. The size of Cooper pairs $\xi$ (coherence lengths) is also suppressed in nano-sized superconductors \cite{Michotte}, and hence it would allow vortices to enter nano-superconductors more easily. Moreover, other studies \cite{Prozorov1} discovered that the magnetisation response depends on the geometry and shape of the samples, and also concluded that geometric barriers are the key for the vortex-pattern. Recently, Zolotavin {\it et al.} \cite{Zolotavin1,Zolotavin2} found that the critical temperature $T_c$ and the critical field $H_c$ could also be altered at the nano-scale, due to the change of electron-phonon interaction and electronic density. 

The vortex states of a nano-superconductor can be logic states in magnetic switching devices, controlled by the strength of an applied magnetic field \cite{Fomin,Forrester1,Forrester2}. There are several innovative proposals for the applications of nano-superconductors \cite{Patrick,Forrester1}, such as Josephson cylinders \cite{Jung,Thurmer}, mixed superconductor-semiconductor devices \cite{Mourik}, and some superconducting nano-particles \cite{Zolotavin1,Zolotavin2,Lain,Geigenmuller,Forrester1,Forrester2}.  

The objective of this article is to understand the behaviours of confined vortices in Type II nano-superconducting devices. We look at the magnetisation states ($1^{st}$ or $2^{nd}$ order phase transition) and vortex structures in nano-superconductors with cylindrical geometries \cite{Baelus1,Baelus2,Baelus3,Antonio,Wu1}, different sizes (from one coherent length $\xi$ to a few) and finally the effects of the orientation of applied magnetic fields (transverse and longitudinal). We also look at the possibility of the giant vortex ({\i.e.} a multi-quanta vortex) in some specific orientations and geometries. Indeed, a giant vortex needs more energy to nucleate than a multitude of vortices \cite{Grigorenko}. All these characteristics are important for the control of flux in nano-devices.

\section{Nonlinear Ginzburg-Landau Model for Nano-Superconducting Devices}
To model our systems, we use the pair of nonlinear coupled Ginzburg-Landau equations (see for example, Abrikosov~\cite{Abrikosov} and Ginzburg~\cite{Ginzburg})
\begin{equation}
\alpha(T){\Psi}+\beta(T){\Psi}|{\Psi}|^2  + \frac{1}{2m^*} \left( \frac{\hbar}{i}\nabla - \frac{e^*}{c}{\bf A} \right)^2 \Psi = 0
\label{gl_1}
\end{equation}
\begin{equation}
{\bf J} = \frac{e^* \hbar}{m^*}{\rm Im}({\Psi^*}\nabla {\Psi}) - \frac{e^{*2}}{m^*c}{\bf A}\,|{\Psi}|^2,
\label{gl_2}
\end{equation}
where ${\Psi}$ is an order parameter, ${\bf A}$ is the magnetic potential (${\bf B} = \nabla \times {\bf A}$), and ${\bf B}$ is the local magnetic field. $\alpha(T)$ and $\beta(T)$ are temperature dependent parameters, near the critical temperature $T_c$, characterized by the superconducting materials. $\nabla = (\partial_x, \partial_y, \partial_z)$ is the gradient operator, and $e^*$ and $m^*$ are the effective charge and mass of the Cooper pair respectively. ${\bf J}$ is the superconducting current density \cite{Gennes1,Gennes2,Abrikosov2,Tinkham,James}.
 
In order to study the superconducting cylinders, in transverse (${\bf H} = H{\bf \hat{x}}$) and longitudinal (${\bf H} = H{\bf \hat{z}}$) applied magnetic fields, we need to solve Eqs.~(\ref{gl_1}) and (\ref{gl_2}), subject to the boundary conditions:
(a) that the normal component of the super-current vanishes (~\cite{Gennes2,Abrikosov2}), $[(\hbar/i)\nabla - (e^*/c){\bf A}]\Psi \cdot {\bf \hat{n}} = 0$
and, (b) that the field strength on the surface $\partial \Omega$ is equal to the applied magnetic field, 
${\bf H}= \nabla \times  {{\bf A}_{\rm app}}$,
where ${{\bf A}_{\rm app}}$ corresponds to the applied magnetic potential on the boundary.
The order parameter ${\Psi}$ and the vector potential $\bf A$ can be calculated by solving the self-consistent Ginzburg-Landau equations.
In the nano-sized scale, the boundary conditions have a significant contribution to the magnetisation response and to the vortex-vortex interactions.
Once the magnetic potential ${\bf A}$ in the system is solved, the average magnetisation can be estimated from
${\bf M} = \frac{1}{4\pi V}\int(\nabla \times  {\bf A} - {\bf H})\,{\rm d}V$, where $V$ is the volume of the system.

\section{Analysis and Studies of Vortices in Confined Geometries}
The Type II superconductors $\kappa = \lambda(T)/\xi(T) > 1/\sqrt{2}$ have been investigated in this study, where $\lambda(T) = \sqrt{m^*c^2/(4 \pi e^{*2} |\Psi(T)|^2)}$ is the London penetration depth and $\xi(T) = \sqrt{\hbar^2/(2m^*|\alpha(T)|)}$ is the coherence length (Cooper pair). Furthermore, in the Ginzburg-Landau model, penetration depth and coherence length can be approximated in terms of: $\xi(T)\approx\xi_0/\sqrt{1-(T/T_c)}$ and $\lambda(T)\approx\lambda_0/\sqrt{1-(T/T_c)}$ respectively, where $\xi_0$ and $\lambda_0$ are the characteristic parameters at $T=0\, \mathrm{K}$. Hence $\kappa=\xi_0/\lambda_0$ which is independent of temperature. Here we choose $\kappa=2>\sqrt{2}$. The superconducting nano-cylinders are in transverse $({\bf H} = H \bf{x})$ and longitudinal $({\bf H} = H \bf{z})$ applied magnetic fields. The dimensions of the nano-devices range from 1$\xi$ to 10$\xi$.

\subsection{Nano-Cylinders with radius $R = \xi$ and length $L = 10 \xi$}
\subsubsection{Reversible Second Order Phase Transition}
The superconducting nano-cylinder ($\kappa = 2$) we study has radius $R = \xi$ and  length $L = 10 \xi$, and in this subsection the applied magnetic field is ${\bf H} = H {\hat x}$ and is transverse to the axis of a nano-cylinder (Fig.1).

\begin{figure}
\begin{center}
\includegraphics[width=8cm,height=5cm]{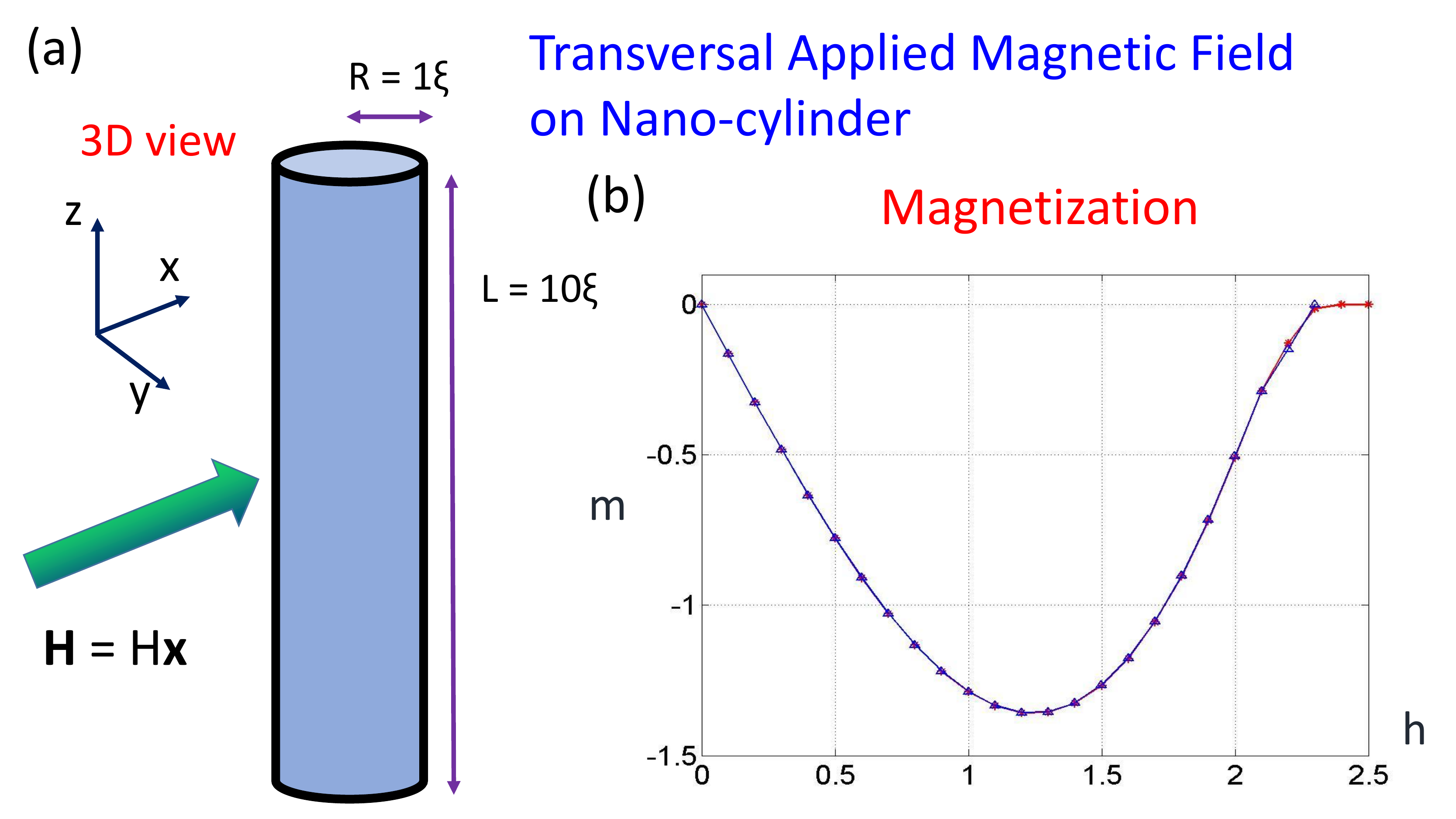} 
\caption{\label{cy_x1} A superconducting nano-cylinder with radius $R = \xi$ and length $L = 10 \xi$. Type II $(\kappa = 2)$ diagrams are shown. Applied magnetic field ${\bf H} = H \bf{x}$ and the dimensionless field $h$ are along the x-axis. The magnetisation curves along with the increasing (red) and decreasing (blue) applied fields are shown. Both curves coincide with the increasing and decreasing fields - which reveals the second order phase transition.} 
\end{center}
\end{figure}

Our dimensionless magnetisation $m=M/H_{c2}$ is as a function of the dimensionless applied field $h=H/H_{c2}$, where $H_{c2}$ is a second critical field of Type II superconductor from the Ginzburg-Landau model, and is defined as
\begin{equation}
\label{Hc2}
H_{c2}=\frac{\phi_{0}}{2 \pi \xi^{2}}.
\end{equation}
Henceforth $h$ denotes the dimensionless applied magnetic field (NOT the Planck's constant). Fig.1 (b) is the magnetisation curves with respect to applied field. The red (star) and blue (triangle) curves are respectively the magnetisation $m$ in an increasing magnetic field and in a decreasing magnetic field $h$ (Fig.1). The magnetisation curve is reversible and exhibits typical second order transition with no clear cut between the Meissner state and normal state as $h$ is varied. Given that the radius of the cylinder is equal to one coherence length, flux cannot be trapped in the cylinder and hence no vortex can nucleate in the system (there is no jump in $m$ in Fig. \ref{cy_x1}). This result is similar to the experimental results of Geim {\it et al.} \cite{Geim1} on superconducting disks of size $R=\xi$. 

In macroscopic superconductors, the relationship between the second critical field $H_{c2}$ and the surface nucleation field $H_{c3}$ can be approximated to (see references~\cite{Gennes1,Strongin,Peeter1})
\begin{equation}
H_{c3}\approx 1.69H_{c2}.
\end{equation}
For the slim nano-cylinder ($R=\xi$) being studied here, $H_{c3}$ is about 2.2 $H_{c2}$. Even though the magnitude of the magnetisation is small, the results show that the cylinder can resist magnetic fields larger than those estimated by de Gennes \cite{Gennes2}. This may possibly be due to both the size, and the cylindrical shape of the sample. The effective coherence length $\xi_{eff}$ is de facto smaller than in theory, and hence $H_{c2}=\phi_{0}/(2 \pi \xi_{eff}^{2})$ is effectively larger than in theory.    

\begin{figure}
\begin{center}
\includegraphics[width=8cm,height=5cm]{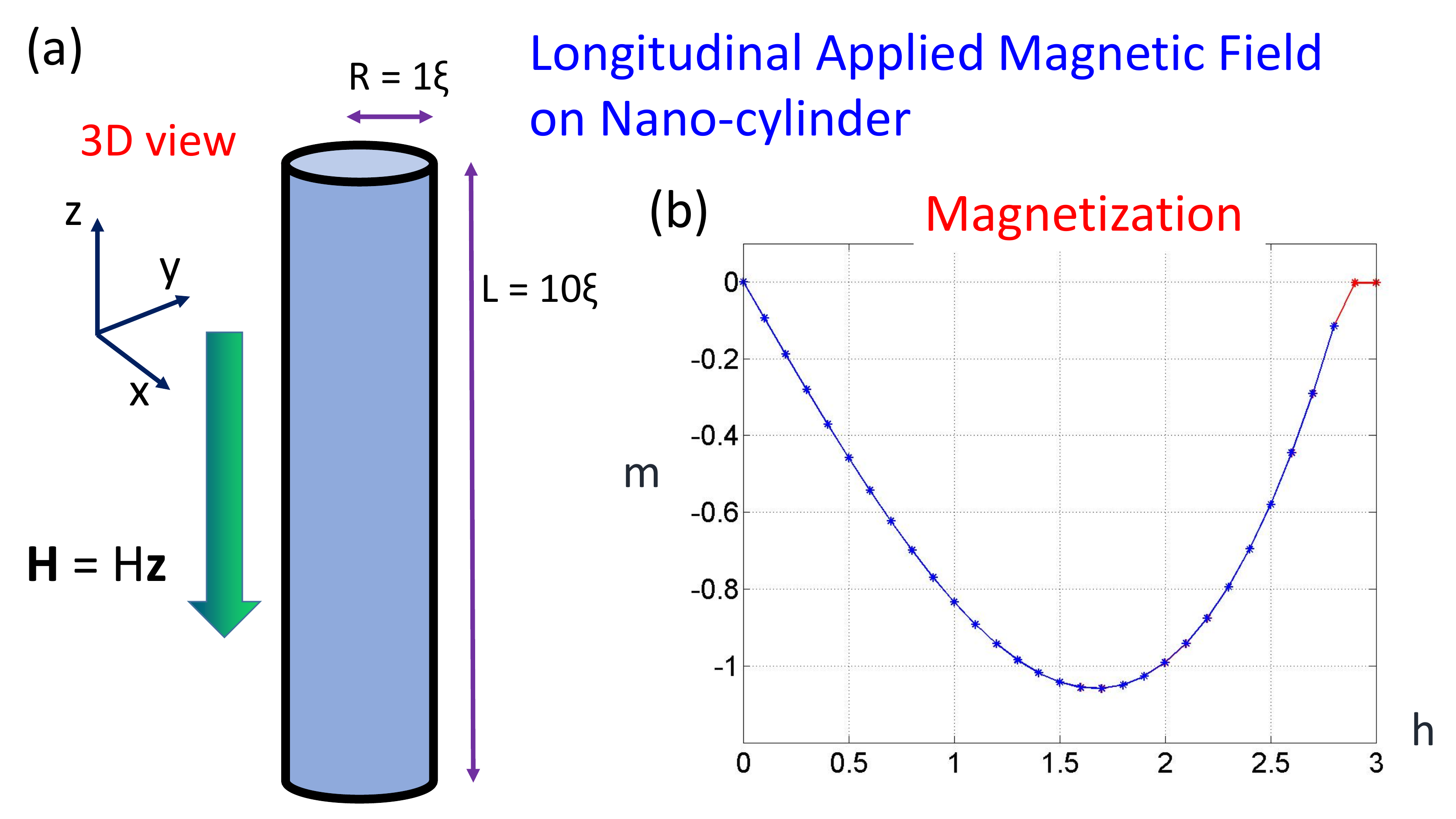} 
\caption{\label{cy_z1} Type II ($\kappa = 2$) superconducting nano-cylinder ($R=\xi$ and $L=10\xi$) in a longitudinal magnetic field. The second order transition is found in the magnetisation, as in the longitudinal magnetic field.}
\end{center}
\end{figure}

We also give the results for superconducting nano-cylinders in a longitudinal applied field ${\bf H} = H {\hat z}$,  parallel to the axis of the cylinders (Fig. \ref{cy_z1}). 
For a cylinder with dimensions $R = \xi$ and $L = 10 \xi$ (with $\kappa =2$), it is found that no vortex is trapped in the cylinder, and the magnetisation exhibits reversible second order transition, just as in the case of transverse fields. However $H_{c3} \sim 2.7 H_{c2}$ which is larger than in transverse fields. It is again
the effective coherence length $\xi_{eff}$ in the simulation is less than in theory, and the $\xi_{eff}$ in longitudinal fields that is smaller than in transverse fields. Therefore, $H_{c3}$ in transverse fields is larger than in longitudinal fields.

\subsection{Nano-Cylinders with radius $R = 3\xi$ and length $L = 10 \xi$}
\subsubsection{Hysteresis on Magnetisation} 

\begin{figure}
\begin{center}
\includegraphics[width=8cm,height=5cm]{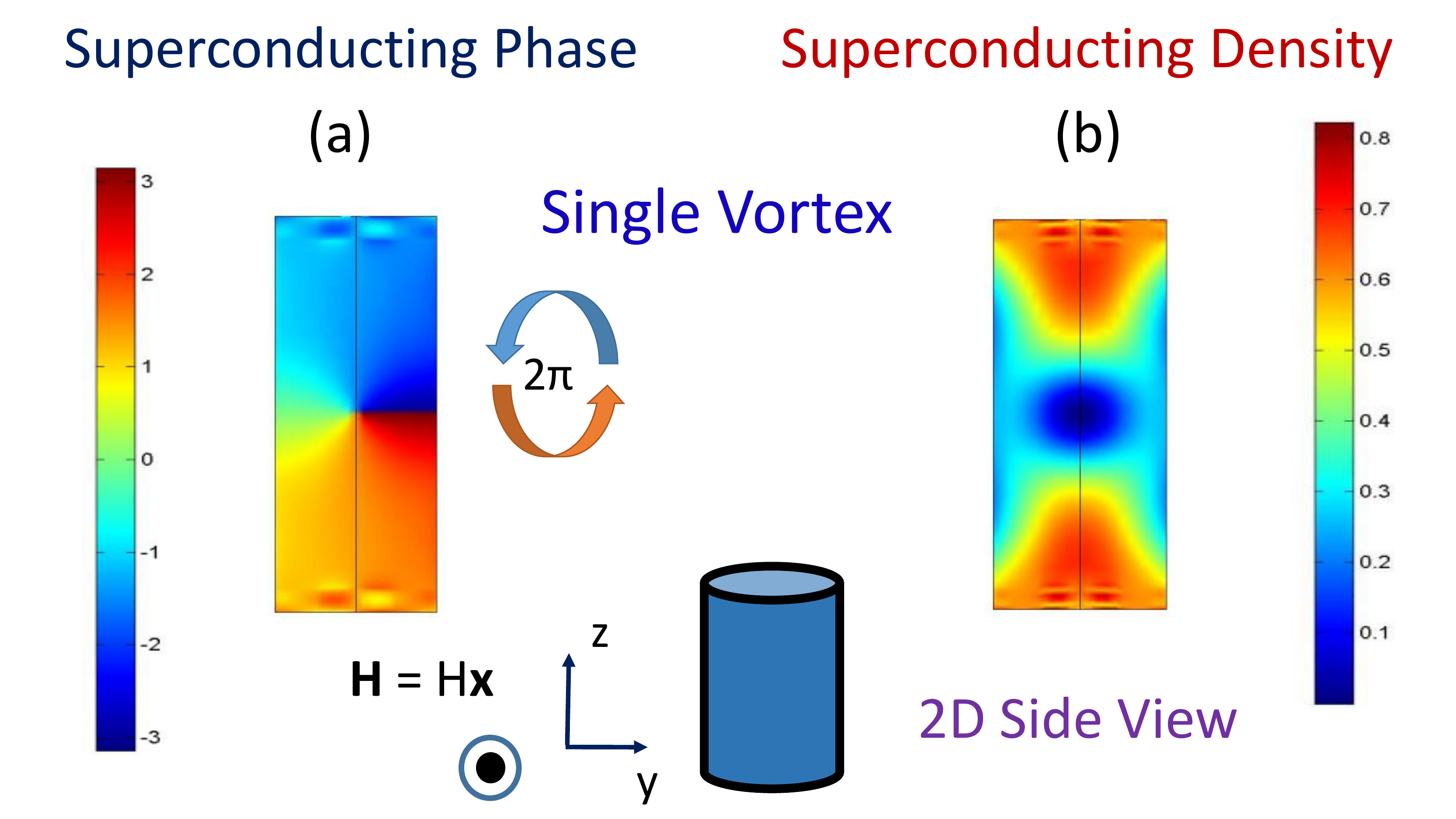} 
\caption{\label{cy_x3_2D}  A nano-cylinder (length $L = 10 \xi$ and radius $R = 3\xi$ $(\kappa = 2)$) is in the transverse magnetic field. The side view (y-z plane) of the nano-cylinder with one vortex is shown. The left diagram (a) is the superconducting phase of the nano-cylinder ($\phi = \operatorname{Im} (\ln {\psi})$). The left colour bar ranges from $-\pi$ (blue colour) to $\pi$ (red colour), there is a phase change of $2\pi$ and represents an integral quantised vortex. The right diagram (b) represents the superconducting density ($|\tilde\psi| = |\psi|/\sqrt{|\alpha/\beta|}$). The blue and red colours (right colour bar) represent low and high superconducting densities, respectively.}
\end{center}
\end{figure} 

\begin{figure}
\begin{center}
\includegraphics[width=9cm,height=6cm]{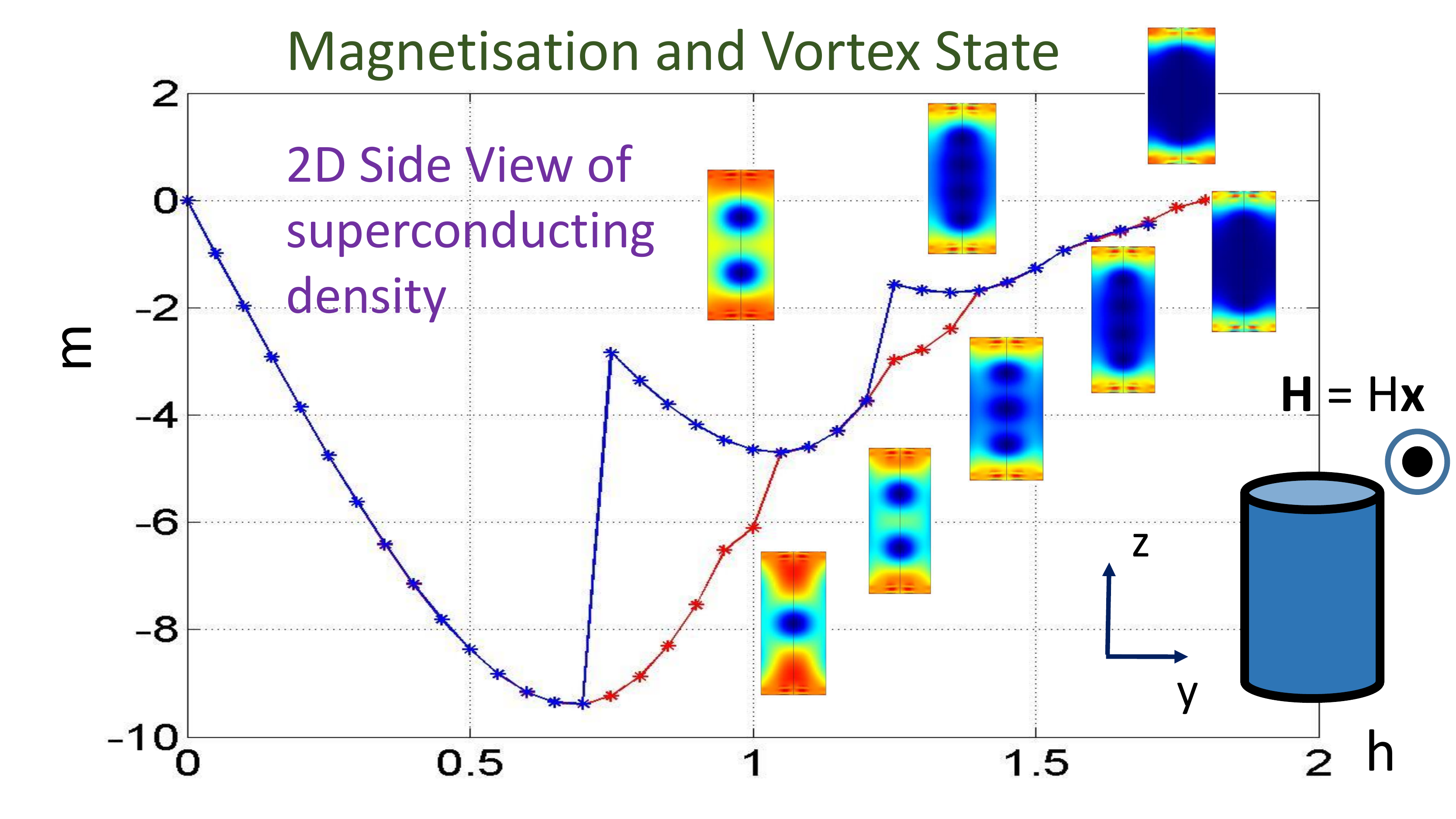} 
\caption{\label{cy_x3} A superconducting nano-cylinder has the length $L = 10 \xi$ and radius $R = 3\xi$ $(\kappa = 2)$. The red (blue) curve is the magnetisation along with the increasing (decreasing) field. The magnetisation $m$ shows obvious hysteresis; this may be due to the surface barrier of the system. Vortices are found to be trapped inside the nano-cylinder.}
\end{center}
\end{figure}

\begin{figure}
\begin{center}
\includegraphics[width=8cm,height=5cm]{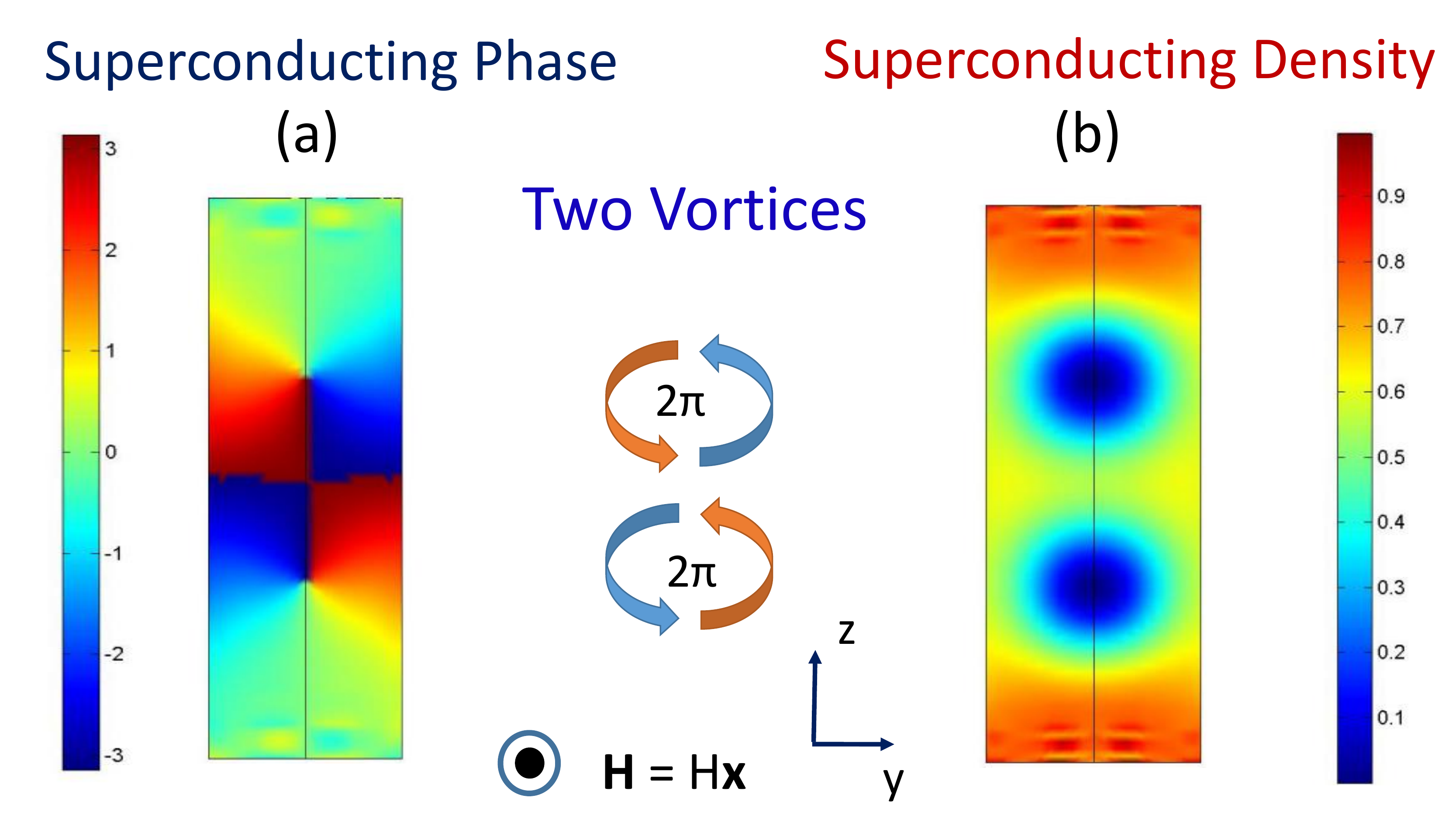} 
\caption{\label{cy_x3_2D_2}  The side view (y-z plane) of the two and four vortices states, in the nano-cylinder. Left diagram (a) is the phase of the nano-cylinder ($\phi = \operatorname{Im} (\ln {\psi})$). From $-\pi$ (blue colour) to $\pi$ (red colour), there are phase change of two `$2\pi$' and represent two integral quantised vortices. Right diagrams (b) represents the superconducting density. Blue and red represent the low and high superconducting density respectively.}
\end{center}
\end{figure} 

Now we consider a nano-cylinder of radius $R = 3 \xi$ and length $L = 10 \xi$ in a transverse applied magnetic field, and study the vortex-pattern. The `diagram' obtained from the simulations will be discussed before going into the results. The side view (y-z plane) of the nano-cylinder with one vortex is shown (Fig.~\ref{cy_x3_2D}). Fig.~\ref{cy_x3_2D}(a) (left diagram) is the superconducting phase of the nano-cylinder ($\phi = \operatorname{Im} (\ln {\psi})$). The left colour bar ranges from $-\pi$ (blue colour) to $\pi$ (red colour). There is a phase change of $2\pi$ that represents an integral quantised vortex. Fig.~\ref{cy_x3_2D}(b) (right diagram) shows the dimensionless superconducting density profiles $|\tilde\psi|^2$ ($|\tilde\psi| = |\psi|/\sqrt{|\alpha/\beta|}$). The blue and red colours (in the colour bar) correspond to low and high superconducting densities respectively. The deep blue circle represents one vortex in the middle of the nano-cylinder.

The magnetisation curves with vortices trapped are shown in the Fig.~\ref{cy_x3_2D} and Fig.~\ref{cy_x3}. The jumps in the magnetisation $m$ (Fig.~\ref{cy_x3}) are signatures of flux being trapped into, or expelled out of, the cylinder. The different vortex states of the system are shown in the density profile plots, where the normalised superconducting density $|\tilde\psi|^2$ along the cylinder is presented (as discussed in Fig.~\ref{cy_x3}). The radius of a vortex is of the order of $\xi$, and hence vortices can nucleate in the cylinder. The magnetisation $m$ shows obvious hysteresis; this may be due to the curved surface (of the cylinder). 

In an increasing magnetic field, the magnetisation of the nano-cylinder shows that the successive few vortex states are: a single vortex state, followed by two-vortex, three-vortex and four-vortex states; as the field is increased further the cylinder reaches the normal state (Fig.~\ref{cy_x3}). In a decreasing field, the first vortex state has five vortices; as the field is reduced, one vortex is expelled, followed by a further two. Reducing the field further drives the system into the Meissner state. The average magnetisation in the magnetic cooling is weaker than in magnetic heating. The reason is that the magnetic field is trapped inside the cylinder in the decreasing field, and cannot be easily expelled from the cylinder because of the surface barrier.

The geometry of the system can possibly change the electron mean free path $l$ near the surface. If $l$ is finite, Pippard~\cite{Gennes2,Abrikosov2,Tinkham,James} showed that the coherence length $\xi$ is actually shorter than in a pure bulk sample, and can be deduced from ${1}/{\xi} = {1}/{\xi_{\rm pure}} + {1}/{l}.$ This effectively results in an increase of $\kappa = \lambda/\xi$ and hence a change in the surface energy between the superconducting and normal phases; this will change the vortex configurations in the nano-cylinders. Similar to Fig. \ref{cy_x3_2D} for one vortex, Fig. \ref{cy_x3_2D_2} shows the superconducting phase $\phi = \operatorname{Im} (\ln {\psi})$ (left) and superconducting density (right) when two vortices are present in the nano-cylinder.

\subsubsection{Giant Vortex in a Confined System}

\begin{figure}
\begin{center}
\includegraphics[width=9cm,height=5.5cm]{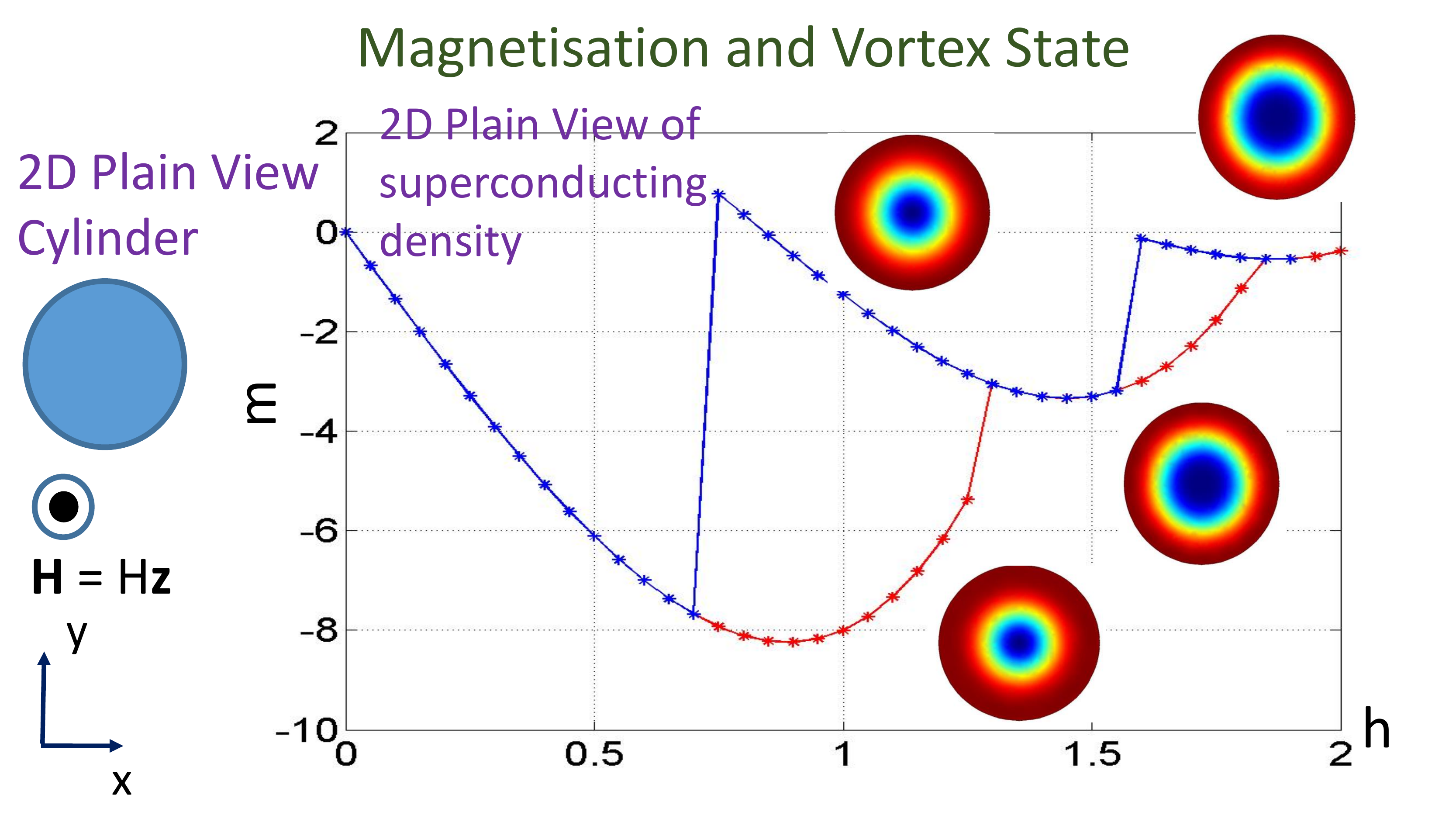} 
\caption{\label{cy_z3}  Type II ($\kappa = 2$) superconducting nano-cylinder ($R=3\xi$ and $L=10\xi$) in a longitudinal magnetic field. The magnetisation exhibits the hysteresis in the increasing and decreasing magnetic fields. We found that the giant vortex has been trapped inside the cylinder. In fact, the radius of the cylinder is not large enough to allow a separated vortex.}
\end{center}
\end{figure}

\begin{figure}
\begin{center}
\includegraphics[width=8cm,height=5cm]{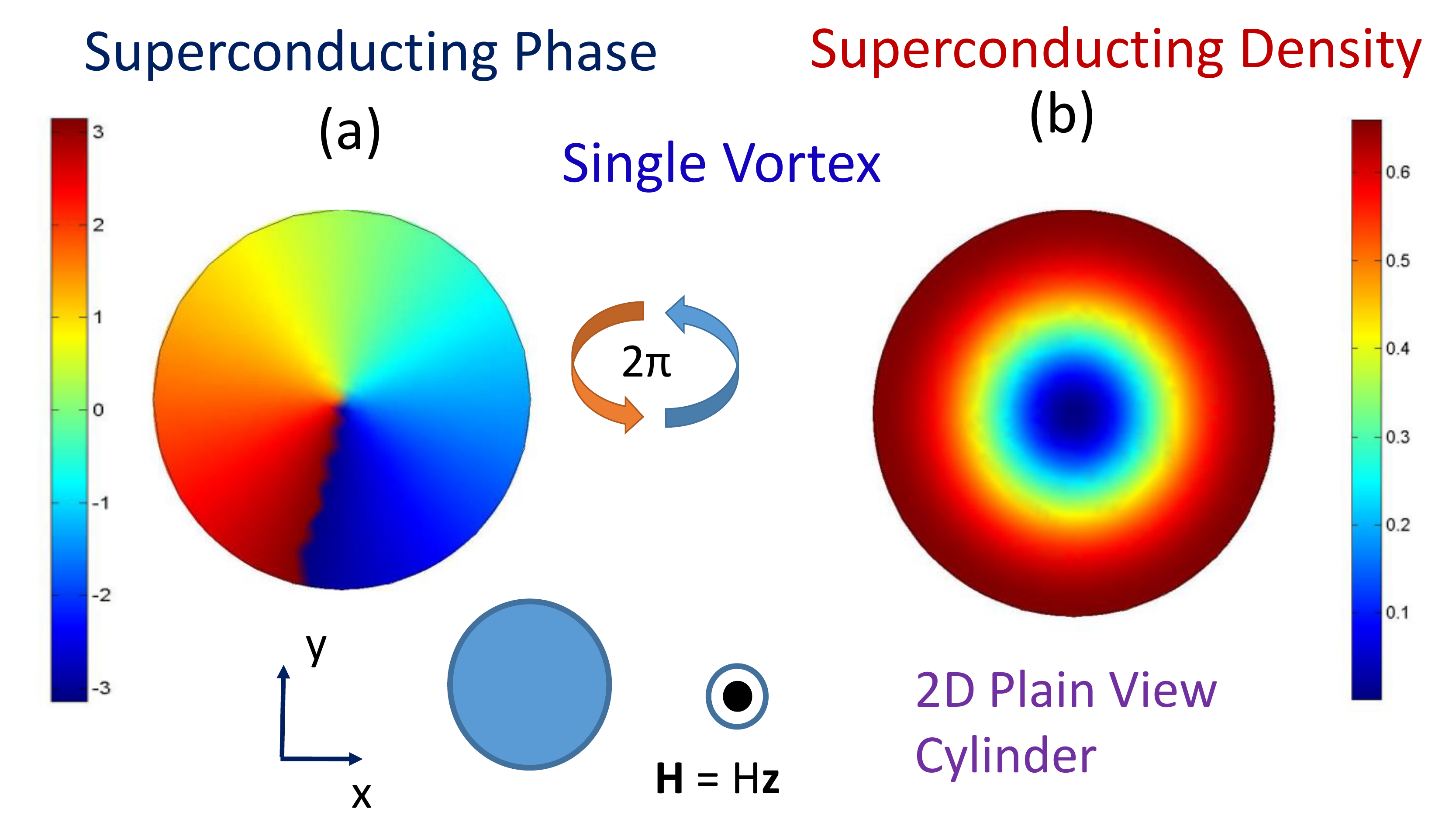} 
\caption{\label{cy_z3_2D} The diagram shows the a nano-cylinder with 2D plain view (x-y plane) in a longitudinal field. The left diagram (a) is the superconducting phase ($\phi = \operatorname{Im} (\ln {\psi})$) of the system. It shows that there is a phase change of $2\pi$ and represents a single quantised vortex. The right diagram (b) represents the superconducting density ($|\tilde\psi|$). The blue and red colours (right colour bar) represent low and high superconducting densities respectively. It can be seen that a vortex is pinned into the centre of the system.}
\end{center}
\end{figure}

Fig. \ref{cy_z3} shows the magnetisation $m$ as a function of $h$ for a cylinder with $R = 3 \xi$ and $L= 10 \xi$ ($\kappa =2$), with longitudinal field $({\bf H} = H \bf{z})$. The magnetisation shows an obvious hysteresis. Unlike the small cylinder ($R=\xi$), vortices can be trapped in the larger cylinder. In an increasing magnetic field $h$, the system evolves from the Meissner state to a state with vortices, and then to the normal state as expected for a Type II superconductor. We identify two different vortex states in an increasing field. The singly quantised vortex state is plotted in the Fig. \ref{cy_z3_2D}. 
The diagram shows the nano-cylinder with 2D plain view (x-y plane) in longitudinal field. Fig. \ref{cy_z3_2D}(a) is the superconducting phase $\phi$ which shows that there is a phase change of $2\pi$. Fig. \ref{cy_z3_2D}(b) represents the superconducting density $|\tilde\psi|$. Both Fig. \ref{cy_z3_2D}(a) and (b) indicate that only a single vortex is pinned into the centre of the system.

\begin{figure}
\begin{center}
\includegraphics[width=8cm,height=5cm]{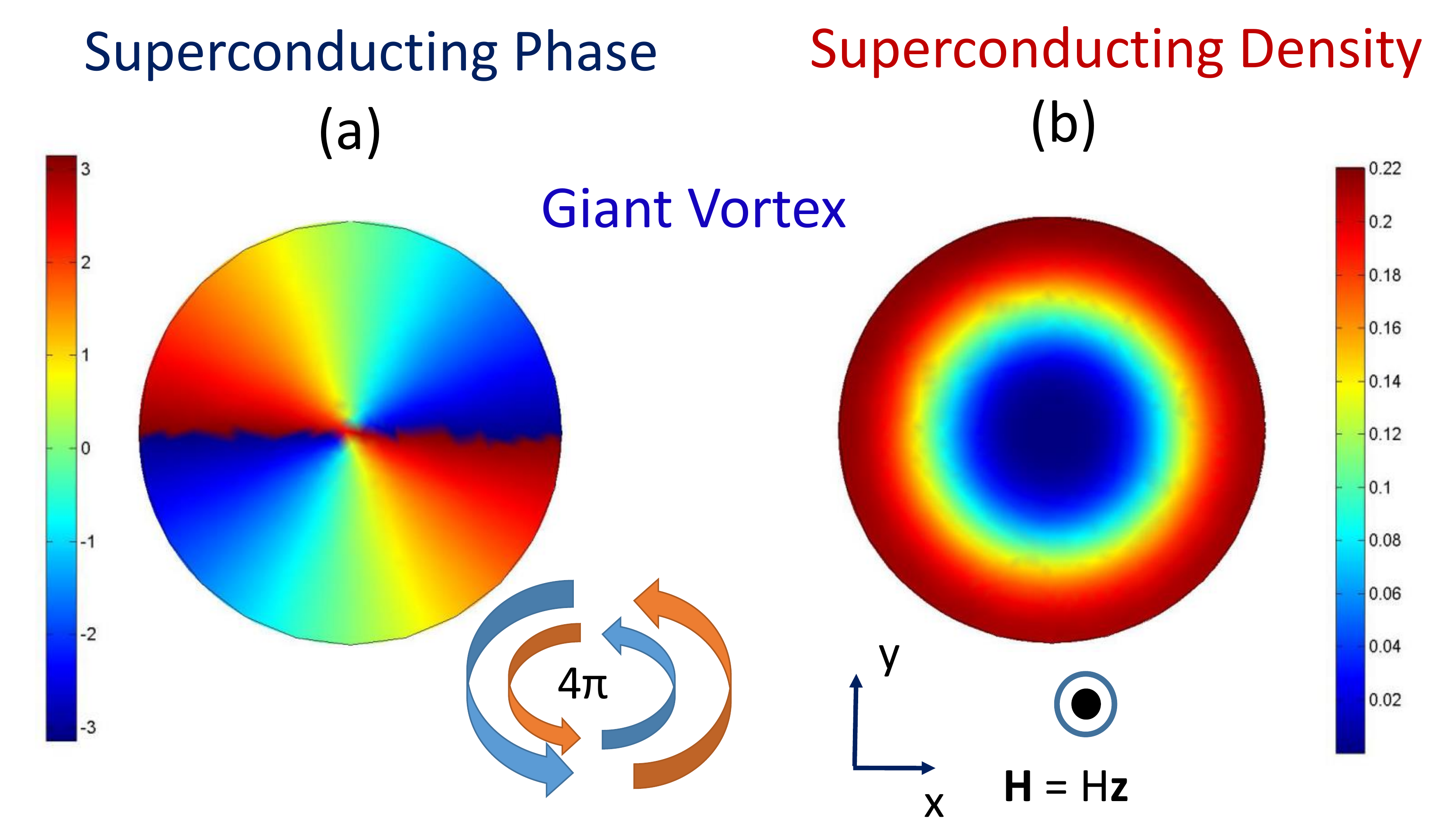} 
\caption{\label{cy_z3_2D_2} This diagram has a similar configuration to Fig. \ref{cy_z3_2D}. The superconducting phase (left diagram) shows that there is a phase change of $4\pi$ and represents a doubly quantised vortex. The superconducting density (right diagram) reveals a vortex state pinning into the centre, and the size of this double fluxoid vortex is obviously larger than a single vortex one shown in Fig. \ref{cy_z3_2D}.}
\end{center}
\end{figure}

Due to the size constraint (the radius is only $3 \xi$), multiple single vortices cannot nucleate. 
Instead a giant, multi-quanta, vortex is formed - see the phase plot of Fig. \ref{cy_z3_2D_2} which shows an obvious phase change of $4\pi$. The superconducting density (right diagram) reveals a vortex state pinning into the centre, and the size of this double fluxoid vortex is obviously larger than a single vortex one shown in Fig. \ref{cy_z3_2D}.
In a decreasing field, the cylinder evolves from the normal state to a state with a four-quanta vortex. As the field is decreased further, flux is expelled from the cylinder and the subsequent vortex states have successively a two-quanta vortex and a Meissner state. The para-Meissner state is also observed in the decreasing field (\cite{Geim1,Geim2,Geim3}). It is due to the flux being trapped inside the system, and it can not be expelled even when the magnetic field decreases (Fig. \ref{cy_z3}).

In general, the vortex configuration (or pattern) is determined by relative sizes of the self-nucleation energy $E_s$, the interaction energy $E_I$ between vortices, and the interaction energy $E_b$ between the vortices and the boundaries.
The energy to nucleate a single 3D vortex is approximately \cite{Tinkham,James}
\begin{equation}
E_s \approx \frac{\phi_{0}^2 C}{16 \pi^2 \lambda^2} \ln\left(\frac{\lambda}{\xi}\right),
\end{equation}
where $C$ is the length of a 3D vortex tube. The interaction energy between two vortices is  
$E_I \approx \frac{\phi_{0}^2 C}{8 \pi^2 \lambda^2} \ln\left(\frac{d_1}{\xi}\right)$,
where $d_1$ is the separation of the vortices \cite{Tinkham,James}.
The interaction energy of a vortex with a boundary (the interaction of a single vortex with the boundary can be described by the method of images with the interaction of a vortex with its image anti-vortex in analogy to electrostatics) is 
$E_b \approx -\frac{\phi_{0}^2 C}{8 \pi^2 \lambda^2} \ln\left(\frac{d_2}{\xi}\right)$, where $d_2$ is in the distance between a vortex and its image anti-vortex and depends on the geometry \cite{Tinkham}. The vortex patterns are determined by the relative sizes of these attractive and repulsive terms ~\cite{Thomas,Novoselov}. The parameters $C$, $d_1$ and $d_2$ are functions of the size and geometry of a sample. Vortex-vortex interactions are usually repulsive, but in this nano-cylinder there is a short range attraction~\cite{Thomas}. The interaction becomes more complicated when there are more than two vortices \cite{Forrester1,Wu1,Forrester2}. That is why some novel hysteric behaviours, associated with first and second order phase transitions occur. Furthermore, the confined geometry gives rise to the possibility of a giant vortex.

\section{Conclusions}
We have studied the magnetisation curves of different Type II superconducting nano-geometries which are of dimensions comparable to the length scales of Cooper pairs and $\kappa = 2$, in transverse and longitudinal applied magnetic fields. Here the magnetisation has been investigated as a function of the applied magnetic field, demonstrating first (reversible) and second (flux trapped and expelled) order transitions of the sizes $R = \xi$ and $R = 3\xi$ respectively. It is due to the fact that the size of $R = \xi$ is too small to allow a vortex. Only the size $R = 3\xi$
 is large enough to form some vortex states. Fluxoids nucleate or are trapped in the nano-cylinders ($R = 3\xi$); the trapping is associated with the jumps in the magnetisation curves. These are signatures of quantised vortices nucleating into (as $H$ is increased) or being expelled from (as $H$ is decreased) the nano-cylinders. We have presented density profiles for the stable vortex states of different ranges of $H$. The magnetisation curves are not reversible -- they show hysteresis; this implies that there are several (at least two) metastable vortex states and different states are accessed in an increasing field and in a decreasing field respectively. We also found a giant (doubly quantised) vortex with $4\pi$ phase change in a cylinder ($R = 3\xi$) with the longitudinal applied fields. The reason for a giant vortex is again caused by the suppressed size of the system.




\begin{thebibliography}{99}
\bibitem{Lond1} London, F. {\it Nature} \textbf {1937}, {\it 140}, 793-796.
\bibitem{Lond2} London, F. {\it Nature} \textbf {1937}, {\it 140}, 834-836.
\bibitem{Dorsey} Dorsey, A. T.  {\it Nature} \textbf {2000}, {\it 408}, 783-785.
\bibitem{Bend1} Engbarth, M.; Bending S. J.; Milosevic, M. V. {\it Phys. Rev. B} \textbf {2011}, {\it 83}, 224504, 1-7.
\bibitem{Bend2} Engbarth, M.; Bending, M. V.; Milosevic, S. J.;  Nasirpouri, F.  {\it J. Phys.: Conf. Ser.} \textbf {2009}, {\it 150}, 052048.
\bibitem{Zhang} Zhang, X. Y.; Dai, J. Y. {\it Nanotechnology} \textbf {2004}, {\it 15}, 1166-1168.
\bibitem{Michotte} Michotte, S; Piraux, L; Dubois, S.; Pailloux, F; Stenuit, G.; Govaerts, J. {\it Physica C} \textbf{2002}, {\it 377}, 267-276.
\bibitem{Stenuit} Stenuit, G.; Michotte, S.; Govaerts, J; Piraus, L. {\it Eur. Phys. J. B}  \textbf {2003}, {\it 33}, 103-107.
\bibitem{Geim2} Geim, A. K.; Dubonos, S. V.; Lok, J. G. S.; Henini, M.; Maan, J. C. {\it Nature} \textbf {1998}, {\it 396}, 144-146.
\bibitem{Aliev} Aliev, A. E.; Lee, S. B.; Zakhidov, A. A.; Baughma, R. H.  {\it Physica C} \textbf {2007}, {\it 453}, 15-23.


\bibitem{Gasparovic} Gasparovic, R. F.; McLean, W. L. {\it Phys. Rev. B} \textbf {1970}, {\it 2}, 2519-2524.
\bibitem{Ishii} Ishii, S.;  Sadki, E. S.; Ooi, S.; Ochiai, Y.; Hirata, K. {\it Physcia C},  \textbf {2005}, {\it 426-431}, 268-272.
\bibitem{Bardeen} Bardeen, J. {\it Phys. Rev. Letters} \textbf{1961}, {\it 7}, 162-163.
\bibitem{Geim1} Geim, A. K.; Grigorieva I. V.; Dubonos, S. V.; Lok, J. G. S.; Mann, J. C.; Filippov, A. E.; Peeters, F. M. {\it Nature} \textbf {1997}, {\it 390}, 259-262.
\bibitem{Geim3} Geim, A. K.; Dubonos, S. V.; Grigorieva I. V.; K. S. Novoselov; Peeters, F. M.; Schweigert, V. A. {\it Nature} \textbf {2000}, {\it 407}, 55-57.
\bibitem{Moshchalkov} Moshchalkov, V. V. ; Glelen, L; Strunk, G; Jonckheere, R;  Qiu, X; Haesendonck, C. V.; Bruynseraede. Y. {\it Nature} \textbf {1995}, {\it 373}, 319-322.
\bibitem{Bean} Bean, C. P.; Livingston J. D. {\it Phys. Rev. Letters} \textbf{1964}, {\it 12}, 14-16.
\bibitem{Novoselov} Novoselov, K. S.; Geim, A. K.; Dubonos, S. V;  Hlll, E. W; Grigorieva, I. V.  {\it Nature} \textbf{2012}, {\it 3}, 810 (1-7).
\bibitem{Peeter1} Schweigert, V. A.; Peeters, F. M.  {\it Phys. Rev. B} \textbf {1999}, {\it 60}, 3084-3087.
\bibitem{Strongin} Strongin, M.; Paskin, A.; Schweitzer, D. G.; Kammerer, O. F.; Craig, P. P. {\it Phys. Rev. Lett.} \textbf {1964}, {\it 12}, 442-444.

\bibitem{Thomas} Thomas, L.; Hayashi, M; Moriya, R; Rettner, C.; Parkin, S  {\it Nature} \textbf{2012}, {\it 3}, 810 (1-7).
\bibitem{Prozorov1} Prozorov, R. {\it Phys. Rev. Lett.} \textbf {2007}, {\it 98}, 257001(4).
\bibitem{Zolotavin1} Zolotavin P.; Guyot-Sionnest, P. {\it ACS Nano} \textbf{2010}, {\it 4}, 5599-5608.
\bibitem{Zolotavin2} Zolotavin P.; Guyot-Sionnest, P. {\it ACS Nano} \textbf{2012}, {\it 6}, 8094-8104.
\bibitem{Fomin} Fomin, V. M.; Rezaev, R. O.; Schmidt, O. G. {\it Nano Letters} \textbf{2012}, {\it 12}, 1282-1287.
\bibitem{Forrester1} Forrester,  D. M.; Kusmartsev, F. V. {\it Scientific Reports} \textbf {2016}, {\it 6}, 25084 (doi:10.1038/srep25084).
\bibitem{Forrester2}Forrester,  D. M.; Kurten, K. E.; Kusmartsev, F. V. {\it Sci. Lett. J.} \textbf {2015}, {\it 4}, 133.
\bibitem{Patrick} Patrick, A L. {\it Nature} \textbf{2000}, {\it 406}, 467-468.
\bibitem{Jung} Jung, M. {\it et al.}, {\it Acs Nano} \textbf{2011}, {\it 5}, 3, 2271-2276.
\bibitem{Thurmer} Thurmer, D. J.; Bufon, C. C. B.; Deneke, C.; Schmidt, O. G. {\it Nano Letters} \textbf{2010}, {\it 10}, 3704-3709.


\bibitem{Mourik} Mourik, V; Zuo, K.; Plissard, S. R.; Bakkers, E. P. A. M.; Kouwenhoven, L. P. {\it Science} \textbf{2012}, {\it 336}, 1003-1007.
\bibitem{Geigenmuller} U. Geigenmuller {\it J. Phys. France} \textbf {1988}, {\it 49}, 405-420.
\bibitem{Lain} S. Lain, J.G. Esteve, G. Waysand {\it Eur. Phys. J. A} \textbf {1999}, {\it 4}, 205-208.
\bibitem{Baelus1} Baelus, B. J.; Cabral, L. R. E; Peeters, F. M. {\it Phys. Rev. B} \textbf{2004}, {\it 69}, 064506, 1-12.
\bibitem{Baelus2} B. J. Baelus, D. Sun, and F. M. Peeters {\it Phys. Rev. B} \textbf{2007}, {\it 75}, 174523, 1-11.
\bibitem{Baelus3} B. J. Baelus, S. V. Yampolskii, and F. M. Peeters {\it Phys. Rev. B} \textbf{2001}, {\it 65}, 024510, 1-10.
\bibitem{Antonio} Antonio R. de C. Romaguera, Mauro M. Doria, and F. M. Peeters {\it Phys. Rev. B} \textbf{2007}, {\it 76}, 020505, 1.
\bibitem{Wu1} Wu, W. M.; Sobnack, M. B.; Kusmartsev, F. V. {\it Int. J. of Mod. Phys. B} \textbf {2010}, {\it 24}, 5093-5104.
\bibitem{Grigorenko} Grigorenko, A.; Bending, S.; Tamegai, T; Ooi, S.; Henini, M. {\it Nature} \textbf {2001}, {\it 414}, 728-731.

\bibitem{Abrikosov} Abrikosov, A. A.  {\it Rev. Mod. Phys.} \textbf{2004}, {\it 76}, 975-979.
\bibitem{Ginzburg} Ginzburg, V. L. {\it Rev. Mod. Phys.} \textbf{2004}, {\it 76}, 981-998.
\bibitem{Gennes1} James, D. S.; De Gennes, P. G.  {\it Phys. Lett.} \textbf {1963}, {\it 7}, 306-308.
\bibitem{Gennes2} De Gennes, P. G. {\it Superconductivity of Metals and Alloys} (Perseus Books Publishing), \textbf {1999}.
\bibitem{Abrikosov2} Abrikosov, A. A. {\it Fundamentals of the Theory of Metals} (Elsevier Science Publishers), \textbf {1988}.
\bibitem{Tinkham} Tinkham, M. {\it Introduction to Superconductivity} (Dover Publication, New York), \textbf {1996}.
\bibitem{James} Annett, J. F. {\it Superconductivity, Superfluids and Condensates} (Oxford University Press), \textbf {2004}.



\end{thebibliography}
\end{document}